# Synchronous motion in a devil's stick – variations on a theme by Kapitza

**Shayak Bhattacharjee**

Department of Physics, Indian Institute of Technology Kanpur, NH-91, Kalyanpur, Kanpur – 208016, Uttar Pradesh, India.

shayak@iitk.ac.in

* * * * *

## Classification

## Abstract

The counter-intuitive rotational motion of the propeller of a devil's stick when the agitator is rubbed against the pylon has long intrigued performers, audiences and scientists. The apparently unrelated phenomenon of self-stabilization of a Kapitza pendulum at the inverted position, once again an amazing lecture demonstration, has on the other hand been subjected to years of investigation and research. Here we show that the two systems are in fact identical, and obtain a previously unreported friction-stabilized synchronously rotating state in the Kaptiza pendulum, which in fact explains the rotational motion of the devil's stick.

* * * * *

## Tema: Kapitza's Pendulum

The problem of a pendulum with vibrating point of support (hereafter referred to as the base) has long been a source of fascination to physicists and engineers. Both vibrational and rotational motions of the system have been discussed extensively in literature. Since the latter are of interest here, we mention a number of studies[1-6] where this motion has been analysed. The techniques range from hardcore analysis[5] to numerical studies[3]. A common feature of these analyses is that a perturbative approach has been used. The bob's acceleration due to the pivot vibrations has been treated as small in comparison with the gravitational acceleration.

A different regime of operation can be seen in studies on vibration-induced inversion of pendula. Kapitza[7] first demonstrated the stabilization of an inverted state, and the analysis was systematized by Landau and Lifshitz[8]. Since then, multiple papers have appeared on the phenomenon, among which the study by Blackburn et. al.[9-10] is notable for their extension of the problem to the non-linear case. We only have the briefest of comments to make about this inversion phenomenon, chief among them being an identification of the sizes of the various forces involved. We note that for a pendulum of length $l$ with base vibrated at amplitude $f$ and frequency $\Omega$, the dynamical equation works out to

$$l\frac{d^2\theta}{dt^2} + \left(g + \Omega^2 f \cos\Omega t\right)\sin\theta = 0 \quad , \qquad (1)$$

where $g$ is the acceleration due to gravity. The effective acceleration thus has a constant component ($g$) and a periodic component ($\Omega^2 f \cos\Omega t$). The analysis proceeds assuming that $\Omega^2 \gg g/l$ and $f/l \ll 1$ and the criterion for stabilization in the linear problem works out to

$$\Omega^2 f^2 > 2gl \quad . \tag{2}$$

From this one can see that $\Omega^2 f/g \approx l/f$ i.e. the periodic component of the acceleration is far larger than the constant component. In other words it is the oscillation which is the dominant effect and gravity plays a perturbative role. The relative weight of the periodic potential is only strengthened if we consider the nonlinear problem instead of the linearization. An approach based on simpler mathematics but more direct physical arguments has been proposed by Pippard[11] where he has also mentioned the possibility of a rotating state. Till date however we have never come across a characterization of the rotating state.

It is interesting to work out the behaviour of the rotating states in this limit, i.e. where the effective force is primarily periodic. This is what we do in the present work, taking the extreme case where gravity does not even exist. Before getting into the analytical details we mention a 'performer's trick' for which we have not found any explanation in the literature and hence propose one here.

## Var. I: The Devil's Stick

The devil's stick is shown in Figure 1a below.

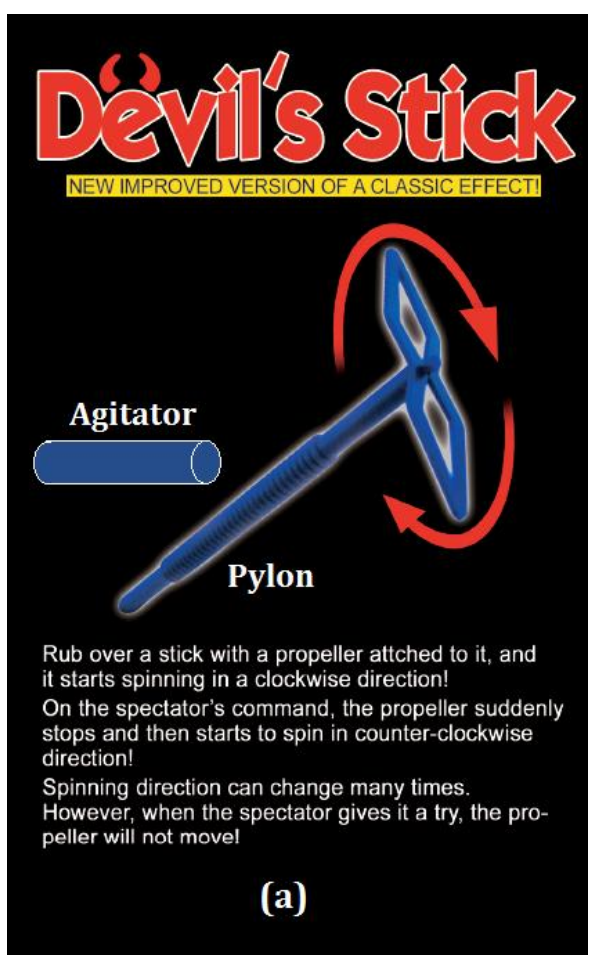

Figure 1 : *The devil's stick, taken from a commercial advertisement ([www.trickproduction.com](www.trickproduction.com)) and slightly modified. The blue stick or pylon is grasped firmly in hand and the agitator is rubbed past the corrugations on the pylon. The propeller is observed to rotate at speed. Note that the propeller is asymmetric with the bottom part larger than the top.*

The variant of the stick considered here consists of a corrugated stick or pylon with a propeller pivoted to it. The propeller is seen to be mounted somewhat eccentrically. Operating it consists of holding the stick firmly in one's hand and rubbing an object (the agitator) such as a pencil up and down the corrugated area. With some practice, it is possible to make the propeller rotate unidirectionally at considerable speed. Let us try analysing the motion of the pylon and hence the pivot when the agitator is rubbed against it. We assume that the pylon is held vertically and the agitator also moves in a vertical direction. Because of the corrugations on the pylon however, the motion of the agitator will be more complex, as seen in the schematic Fig. 2(a) below.

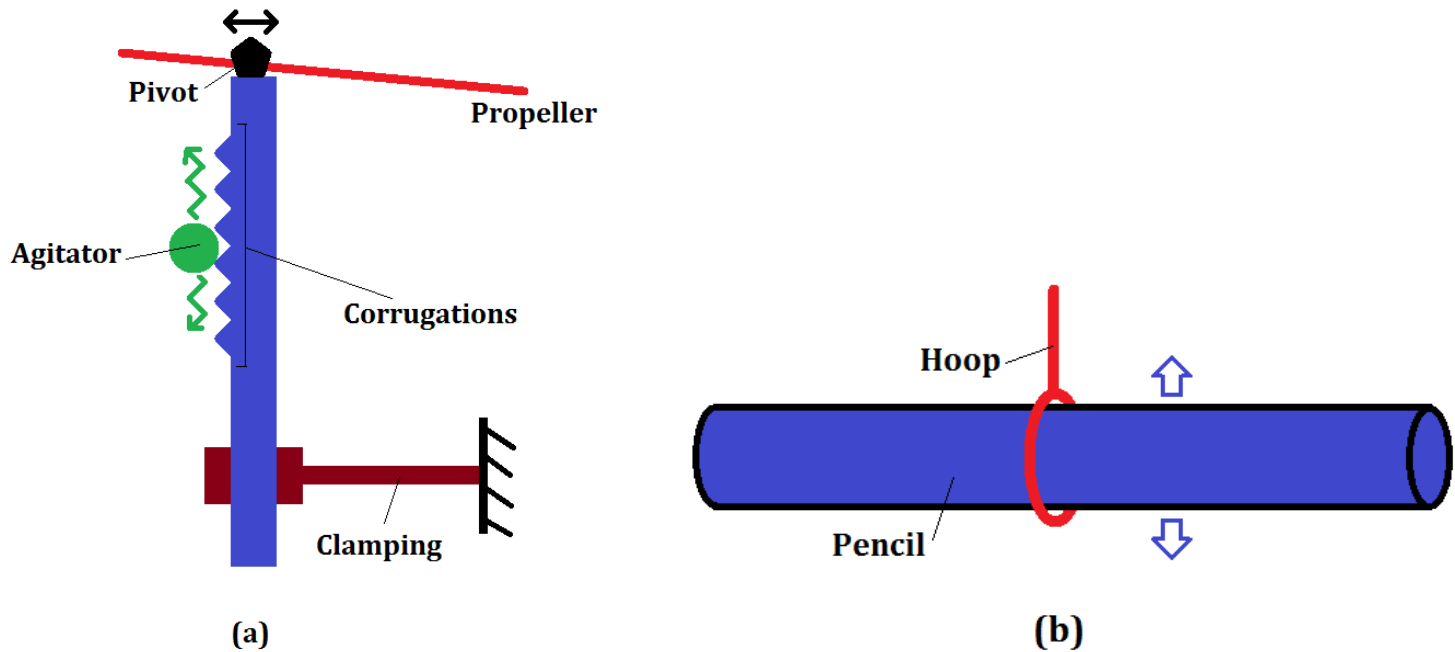

Figure 2 : (a) *Schematic representation of the devil's stick where the motion of the agitator is shown in the green trajectory and the motion of the pivot in the black one. The clamping models the effect of gripping the pylon in hand.* (b) *Variation of the stick where a pendulum with a hoop type attachment is slipped round a vibrating pencil.*

The agitator will have both a vertical and a horizontal oscillatory component to its motion (as seen in the green trace in the figure) and these components will also be imparted to the pylon. Now the effect of holding the pylon in hand will be similar to that of the clamping shown in the schematic and this will eliminate the pylon's vertical motions. The horizontal motion however will not be damped by the clamping and because of this the net motion of the pivot will primarily be a horizontal vibration. We claim that the propeller is excited by this horizontal vibration. Since a Kapitza pendulum features a vibrating pivot, it follows that if there is a rotating state in the stick, there will also be a rotating state in the pendulum. Since gravity plays no role in a configuration where the pylon is vertical, a perturbative treatment with the periodic force much weaker than gravity will not be relevant, and an analysis of the motion in a strong periodic force field will be essential.

Now a quantitative determination of the frequencies of pivot vibration and propeller rotation in the devil's stick is frustrated on two counts : (a) the speed of rotation is very high and is difficult to measure and (b) the effective frequency of horizontal vibration of the pivot will depend in a non-trivial way on the (measurable) frequency of vertical vibration of the pylon. To resolve these issues we repeat the experiment with a simpler setup, shown in Fig. 2b. This consists of an apparatus where a pendulum with a hoop type attachment is slipped round the pencil which is then held in hand and vibrated in the vertical or horizontal plane. If the pencil is thin enough, then any effects arising from its finite size can be neglected and the pivot can be assumed to be at its centre. In this setup, the actual excitation frequency for the pendulum is the same as the frequency of vibration we impart to the pencil, and hence is easily measurable. With some practice, it is once again possible to generate a state where the pendulum rotates and measurement yields the following remarkable relation : *the frequencies of the rotation and vibration are equal*. In other words it is a synchronous state which is found to be stable.

## Var. II: Synchronous Motions in the Pendulum

Synchronization is a diverse yet universal phenomenon,[12-14] occurring in an unimaginable plethora of systems – classical[15,16] and quantum mechanical,[17] electromagnetic,[18] biological and ecological[19,20] as well as behavioural[21]. The techniques used for analysing the system are often not standardised, and a different approach may be required for analysing a different system. Here too we adopt a slightly unconventional track, constructing an intuitive and detailed physical argument to show the existence of a synchronized state, motivate its stability and estimate its basin of attraction.

The first step is to express the oscillating field as a sum of two rotating fields, one clockwise and the other counterclockwise. This manoeuvre is standard in electrical engineering where the single phase magnetic field is broken down into two three-phase fields, and in optics where a linear polarization is expressed as a sum of two circular polarizations. Indeed, a periodic force field in the *x* direction, $2F\cos\Omega t\hat{\mathbf{i}}$, can be expressed in terms of two rotations in the *x-y* plane as $F\left[\left(\cos\Omega t\hat{\mathbf{i}} + \sin\Omega t\hat{\mathbf{j}}\right) + \left(\cos\Omega t\hat{\mathbf{i}} - \sin\Omega t\hat{\mathbf{j}}\right)\right]$. It can readily be verified that the first parenthesis encloses a field which rotates counterclockwise and the second parenthesis contains a clockwise rotating component. The subsequent calculation will focus on the motion of the pendulum through each of these two rotating fields. The dynamic equation from the frame of any one rotating field is

$$I\frac{d^2\delta}{dt^2} + Fl\sin\delta = 0 \quad , \tag{3}$$

where *I* is the moment of inertia of the pendulum, *F* the force on the centre of mass (CM) due to the periodic field, *l* the distance from the pivot to the CM, and *δ* is the torque angle, i.e. the angle between the field and the pendulum. This is of course nothing but the equation of a pendulum in a gravitational field, a system which we will hereafter refer to as "ordinary pendulum". An identical equation arises in the theory of synchronous motors, in which field it is known as swing equation. From the static frame, the same can be written as

$$I\frac{d^2\theta}{dt^2} + Fl\sin\left(\theta - \theta_f\right) = 0 \quad , \tag{4}$$

where *θ* is the usual angle of the pendulum with some arbitrary reference line and $\theta_f$ the angle made by the rotating field with the same line. Of course the oscillating field will have two rotating components, one with $\theta_f = \Omega t$ and the second with $\theta_f = -\Omega t$. Now let us consider a pendulum which has been released with some initial angular velocity *ω* which is quite different from Ω. From the point of view of each rotating component, the pendulum looks like an ordinary pendulum which has been set into rotation at a high speed. (Of course the speeds with respect to the two components are still grossly unequal – we just assume that they are both quite high compared to the pendulum's natural period.) Now such a state is sustainable, with the pendulum showing speed fluctuations over a cycle but maintaining the same speed on the average. The constancy on average will be true as the torque arising from both the field components is rapidly oscillating in time and averages out to zero over any considerable period. It is quite likely that solutions for the fluctuations will be found which will satisfy the equations for both the rotating fields, maintaining the long term average speed *ω*. A more detailed analysis of this situation is however obviated by the realization that, since the average field torque is zero, external torque on the pendulum if any must also average to identically zero over a long time. For a practical problem such as the devil's stick this assumption is fanciful. In reality, there will be a friction term which will act throughout to oppose the motion and thus exert a torque which is negative on the average. This friction, however small, will kill off the rotating state, just as an ordinary pendulum with frictional pivot will pretty soon come to rest irrespective of the release angular velocity. So will the friction eliminate all the rotating states ?

"Save one who stout as Julius Caesar" – *the only state which can maintain itself in the presence of friction is a state where the long time average of the field torque is clearly different from zero*. From (4), that can happen only if the argument of the sine is constant i.e. $\theta = \theta_f + \alpha$ for constant *α* or the state is synchronous. Then the pendulum will appear to make a constant angle *α* with one of the field components and derive a steady torque from it to oppose the friction torque. The other field component will appear to oscillate at twice the frequency and the torque will average out to zero very quickly. The observations with the setup of Fig. 1b are in clearly in accordance with this claim. This picture can also be used to explain the working of the devil's stick.

## Var. III: Stability; Supersynchronous Launch

Qualitative arguments can be used to explain the stability of the synchronous state to minor perturbations. We analyse the situation from the point of view of the rotating field with which the pendulum is synchronous. The other component can be neglected because of its zero average torque. In this synchronous frame, the pendulum appears just like an ordinary gravitational pendulum, a visualization we will frequently resort to in constructing this argument. In a static frame the frictional torque can be assumed to be directly proportional to the angular velocity so in the synchronous frame the friction will be manifest as a constant torque over and above the field torque. Thus the equilibrium position of the equivalent ordinary pendulum makes a finite angle with the vertical. Small disturbances merely cause the pendulum to oscillate about this mean position, maintaining the average speed over the whole cycle. We note that this behaviour is quite different from the motions found in Refs. [1-5]. There, the rotation may be synchronous but may also be at a harmonic or subharmonic of the vibration frequency. Here however, the friction ensures that synchronous state is the only sustainable state and any other frequency response is not admissible.

Having motivated the existence and stability of the synchronous state the next step is to obtain the basin of attraction. The calculations are very difficult so certain simplifications will be used, taking care not to rob the analysis of its principal physical features. The initial release conditions will feature two main parameters – the launching angular velocity and the launch time angle made by the pendulum with the oscillating field. The argument of the preceding paragraph suggests that if the synchronous state be attained once, it may be maintained for all time. Let the pendulum somehow acquire an angular velocity of $\Omega$ (it can be $-\Omega$ with no change in physical consequences; the direction of rotation depends on the initial conditions). For the equivalent ordinary pendulum, it is analogous to releasing the bob with zero angular velocity from any position. In the absence of the extra frictional torque, the motion definitely remains bounded always, but when friction enters the picture this will not hold true. For the frictional pendulum there will be a set of positions from where a zero velocity release will also serve to send the pendulum into a rotating state. However, if the friction is small, then the set of problematic release positions will also be small. Previously we have seen that even a very small friction will suffice to eliminate an asynchronously rotating state. We now use that observation to assume that the friction is sufficiently small so that *a synchronous state, once attained, is perpetuated for ever*.

The preceding statement implies at once that if the launch be supersynchronous i.e. if the launch angular velocity be greater than the synchronous speed then the pendulum will eventually settle down in a synchronous state. For at any higher speed, the torque of both field components will average out to zero over some finite time frame and during this time the effect of friction will cause a retardation of the angular velocity. Since angular velocity decreases continuously it must at some point become equal to the synchronous speed and then the state will be perpetuated. Hence *a launch angular velocity greater than the synchronous speed is sufficient to ensure eventual motion in a synchronous state*.

## Var. IV: Subsynchronous Launch

The problem is more interesting when one considers launch angular velocities less than the synchronous speed. Here it is likely that both the release angle and the velocity will play a role in deciding whether the synchronous state is attained or not. Considering (4) we first make the obvious substitution $\theta_f = \pm \Omega t$. Next we assume that the cyclical small-amplitude fluctuations over the average angular velocity of the pendulum can be ignored, so that we can write $\theta = \omega t$. There is no restriction however that the field and the pendulum must be in phase, so the correct expression for the pendulum's position should be $\theta = \omega t + \beta$ for some phase angle $\beta$. Since the bulk of the discussion is in terms of averages, it makes sense to average (4) over one cycle of the external field. This will be possible if $\omega$ does not change appreciably during this period, which we have already assumed as true.

Considering both the rotating field components as well as a friction term $-k\omega$, the average torque over one cycle of the external field evaluates to

$$\Gamma_{avg} = \frac{2Fl\Omega\omega}{\pi(\Omega^2 - \omega^2)} \sin\left(\frac{\pi\omega}{\Omega}\right) \sin\left(\frac{\pi\omega}{\Omega} + \beta\right) - k\omega \quad . \tag{5}$$

Defining $\varepsilon = \Omega - \omega$ and working in the regime where $\varepsilon/\Omega \ll 1$ the above expression simplifies to

$$\Gamma_{avg} = Fl \sin\left(\frac{\pi\omega}{\Omega} + \beta\right) - k\omega \quad . \tag{6}$$

We note that $\varepsilon$ is a very commonly used variable in the theory of electric motors, where it is called the slip frequency, a terminology we adopt here. In the same theory, $\varepsilon/\Omega$ is denoted by *s* and called slip hence in that jargon, (6) can be said to be the small slip limit of (5). We note that for a synchronous state *s* will be identically zero. For a state where the pendulum is stationary, *s* will equal unity.

The behaviour of the phase term involving $\beta$ is the next topic of interest. Over one external cycle, the rotating field comes back to its initial position and the pendulum advances by $2\pi\omega/\Omega$ hence the value of $\beta$ will also advance by $2\pi\omega/\Omega$ with each cycle. At small slips it is more convenient to view the *degradation* in $\beta$ over one cycle; this amounts to $2\pi s$. Also we make a change of variables, using $\varphi$ to denote $\dfrac{\pi\omega}{\Omega}+\beta$. We see that $\varphi$ is in fact a key player in this synchronisation game.

Its initial value $\varphi_0$ is directly related to the angle with which the pendulum is launched into the oscillating field. If it starts off with a negative value, the torque will be negative at the outset and will tend to retard the pendulum further. On the other hand, a positive initial value will create a positive torque which might bring the pendulum upto synchronism. So long as the pendulum is asynchronous, $\varphi$ will continue to move backwards, potentially changing the sign of the torque. The values of $\varphi$ at which this sign change takes place i.e. at which the torque vanishes, are important. One value $\varphi_1$ will be slightly greater than zero while the second value $\varphi_2$ will be close to $\pi$. Assuming the friction to be small, the deviations of $\varphi_1$ and $\varphi_2$ from 0 and $\pi$ are readily seen to be *Ik$\omega$/Fl*. Suppose that at some point the pendulum, not yet in synchronism, has reached a state where $\varphi=\varphi_1$. Then, as the phase continues to degrade, it will go through a band where the torque is negative, followed by a band where torque is positive. But due to the friction, the size of the negative band is larger than that of the positive band; also the torque magnitude is higher in the negative band as the two signs in the right hand side (RHS) of (6) are parallel. Because of this, the overall effect as the phase degrades through $2\pi$ will be to slow the pendulum down, and this will be repeated periodically. Hence if the synchronous state is not attained when $\varphi$ hits the value $\varphi_1$, it will not be attained ever. Moreover the minimum possible launch angular velocity $\omega_{min}$ required for achieving synchronism corresponds to a case where $\varphi_0=\varphi_2$ as the pendulum then gets the entire positive torque band to accelerate to the synchronous speed. *In terms of the speed at $\varphi=\varphi_1$ and $\varphi_2$ we can then predict the outcome (successful or failed synchronism) of any arbitrary launch.* Let the launch time slip $s_0$ (greater than zero) be given; then $\varphi_0$ can either lie in the positive or negative band. In the former case, it is sufficient to track the system's behaviour until $\varphi$ degrades to $\varphi_1$ (if it does so at all) and then see whether *s* reaches the value zero before this. If it does, then synchronism is achieved, else it is not. In the other case of $\varphi_0$ lying in the negative band it suffices to track the behaviour until $\varphi$ degrades to the value $\varphi_2$. Then synchronism will be achieved if and only if the angular velocity at this point satisfies $\omega>\omega_{min}$ or the corresponding slip, $s<1-\dfrac{\omega_{min}}{\Omega}$.

Finally we give a concrete example to show how the method actually works. For convenience we model the $\sin\varphi$ term as a triangular wave so that the function remains piecewise linear and is easier to handle. Taking an example where $\varphi_0$ lies between $\varphi_1$ and $\pi/2$, let *n* denote an index which counts the number of cycles. Then we have

$$\varphi_{n+1} = \varphi_n - 2\pi s_n \quad , \tag{7a}$$

$$s_{n+1} = s_n - 2\pi\left(\frac{2Fl\varphi_n}{I\Omega^2} - \frac{k}{I\Omega}\right) \quad , \tag{7b}$$

where the triangular wave assumption is seen in (7b), which has been obtained by integration of (6) over a cycle and casting the result in terms of slip. This discrete system may be written as the following matrix relationship between the $n^{th}$ and $(n+1)^{th}$ orders of the vector $[\varphi,s,c]^T$ where *c* denotes the constant quantity $2\pi k/I\Omega$ :

$$\begin{bmatrix}\varphi_{n+1}\\s_{n+1}\\c\end{bmatrix} = \begin{bmatrix}1 & -2\pi & 0\\-\dfrac{4\pi Fl}{I\Omega^2} & 1 & 1\\0 & 0 & 1\end{bmatrix}\begin{bmatrix}\varphi_n\\s_n\\c\end{bmatrix} \quad . \tag{8}$$

The above equation is easily solvable, and thus we have reduced the original complex nonlinear problem to a tractable form. By iterating repeatedly one needs to check whether *s* becomes zero before $\varphi$ hits the critical value $\varphi_1$. As we have already mentioned, (8) is valid for the linear piece of $\sin\varphi$ between $-\pi/2$ and $\pi/2$; for other pieces there will be minor differences but the structure will remain unchanged.

## Coda: Discussion

On the basis of our model let us try to explain some of the phenomena mentioned in the commercial of Fig. 1. We note that the basin of attraction of the synchronous state extends from somewhat below to way above the synchronous speed. Using this we can explain why 'spectators' cannot operate the stick. Since achievement of synchronization requires a launch speed at least quite close to the synchronous speed, it thus follows that there can be two ways of starting the system from rest. In the first method one can give the propeller a large initial impulse in the direction in which synchronization is desired and thus get it to near-synchronism from where entrainment can complete. In the second method one can try varying the vibration frequency slowly from zero upto the target frequency so that at each frequency increment, the propeller can remain within the basin of attraction of the updated frequency state. The lay persons are unlikely to be familiar with either of these methods and hence the stick will not start. These starting algorithms also tell us how the direction of rotation will be determined. Since the vibration has both components in equal measure, it is the initial condition which will determine the direction. We can also explain why the stick can change direction suddenly. Even the most expert operator cannot possibly expect to achieve vibrations of the pivot at truly constant frequency. The driving frequency is likely to fluctuate with time. If the frequency decreases there is no problem but a sudden increase can see the propeller out of the basin of attraction, whereby it will come down to rest. Further irregularities in the motion of the pivot can then give the propeller an impulse in either direction whereby the remaining steps to entrainment can be completed. This accounts for the propeller suddenly flipping direction.

We have thus proposed an explanation behind the working of the devil's stick, and in the process obtained a previously unreported synchronously rotating state in the Kapitza pendulum. Yet another system which follows the equations described here is the 2-pole single phase synchronous motor where there are no capacitors or other phase-splitting equipment in the stator. In fact, this apparatus is possibly the most suited for quantitative measurements. The stator flux will play the role of the oscillating field and the rotor will act as the pendulum. An external mechanism can be used to set initial values of $s$ and $\varphi$ and then see whether synchronism is attained. A further variation which is possible with the motor setup is the introduction of amortisseur windings which create a damping torque proportional to the slip. The effects of such windings can be estimated by repeating the above argument *mutatis mutandis*; the most significant feature is that the stability of the synchronous state is enhanced. Also, the periodic motions arising from the synchronised field get eliminated and the system settles down to a constant torque angle balancing the external torque if any.

\* \* \* \* \*

## Acknowledgement

I am grateful to Kishore Vaigyanik Protsahan Yojana (KVPY), Government of India, for a generous Fellowship.